\pdfoutput=1
\documentclass[aps,prb,twocolumn,amssymb,superscriptaddress,showpacs]{revtex4-1}
\usepackage{float}
\usepackage{xcolor}
\usepackage{graphicx}
\usepackage{dcolumn}
\usepackage{bm}
\usepackage{hyperref}
\usepackage{color}
\usepackage{amsmath}
\usepackage{amsfonts}
\usepackage{setspace}
\usepackage{tikz}
\usetikzlibrary{arrows, shapes, chains}
\usepackage{environ}

\begin{document}

\title{Testing density functional theory in a quantum Ising chain}

\author{Jiahao Mao}
\affiliation{State Key Laboratory of Low Dimensional Quantum
Physics, Department of Physics, Tsinghua University, Beijing
100084, China}
\affiliation{Institute for Advanced Study, Tsinghua University,
Beijing 100084, China}

\author{Haifeng Tang}
\affiliation{State Key Laboratory of Low Dimensional Quantum
Physics, Department of Physics, Tsinghua University, Beijing
100084, China}
\affiliation{Institute for Advanced Study, Tsinghua University,
Beijing 100084, China}

\author{Wenhui Duan}
\email{E-mail: duanw@tsinghua.edu.cn}
\affiliation{State Key Laboratory of Low Dimensional Quantum
Physics, Department of Physics, Tsinghua University, Beijing
100084, China}
\affiliation{Institute for Advanced Study, Tsinghua University,
Beijing 100084, China}

\author{Zheng Liu}
\email{E-mail: zheng-liu@tsinghua.edu.cn}
\affiliation{State Key Laboratory of Low Dimensional Quantum
Physics, Department of Physics, Tsinghua University, Beijing
100084, China}
\affiliation{Institute for Advanced Study, Tsinghua University,
Beijing 100084, China}

\begin{abstract}
By using the quantum Ising chain as a test bed and treating the spin polarization along the external transverse field as the ``generalized density'', we examine the performance of different levels of density functional approximations parallel to those widely used for interacting electrons, such as local density approximation (LDA) and generalized gradient approximation (GGA). We show that by adding the lowest-order and nearest-neighbor density variation correction to the simple LDA, a semi-local energy functional in the spirit of GGA is almost exact over a wide range of inhomogeneous density distribution. In addition, the LDA and GGA error structures bear a high level of resemblance to the quantum phase diagram of the system. These results provide insights into the triumph and failure of these approximations in a general context.
\end{abstract}

\maketitle

\section{Introduction}
The density functional theory (DFT) for interacting electrons has become an indispensable computational tool for condensed-matter physics, quantum chemistry, and even molecular biology~\cite{RN228, RN251}. By utilizing electron density, rather than wavefunction, as the central variable, the complexity of the many-body problem is drastically reduced. The Hohenberg-Kohn (H-K) thereom~\cite{RN266} provides a solid theoretical basis for the success of DFT, stating that completely and uniquely the ground state electron density determines the system, and thus all the properties are in principle functionals of density.

The power of H-K theorem is not limited to interacting electrons. Consider a general many-body system in the form of:
\begin{eqnarray} \label{Hgeneral}
H=H_{int}+H_{ext} \nonumber \\
H_{ext}=\sum_i g_i \hat O_i,
\end{eqnarray}
where $H_{int}$ represents a fixed and parameter-free intrinsic part of the system, and $H_{ext}$ represents the coupling between some local operator $\hat O_i$ and an external field $g_i$.  So long as different $\{g_i\}'s$ do not share a common eigenstate, it is straightforward to prove, by using H-K's \textit{reductio ad absurdum}, that the generalized ground-state density $\langle \hat O_i\rangle$ in principle dictates everything about the system~\cite{RN254}.

Despite the formal simplicity, an explicit construction of density functionals of useful physical properties is highly nontrivial. Since the establishment of H-K theorem, a main theme underlying the development of DFT is to search for better approximations of density functionals, in particular for the ground-state energy. Generations of energy functionals have been formulated and extensively examined for interacting electrons~\cite{RN269,  RN262, RN263, RN248}. Typically, the energy functionals are expected to be exact at the homogeneous limit, and then extended to inhomogeneous cases based on some semi-local approximations. 

A fundamental question is how far these semi-local extensions apply, or under what conditions they fail. For interacting electrons,  it is extremely difficult to provide a transparent answer. The inhomogeneous electron gas is known to exhibit rich phases and unusual properties fundamentally different from a Femi liquid commonly seen in the homogeneous electron gas, but a thorough theoretical understanding is still elusive.

In this article, we switch from interacting electrons to interacting spins, and use the latter as a ``theoretical laboratory'' to investigate this question. Specifically, we base our discussion on a quantum Ising chain. The choice has several advantages. Firstly, it has exact solutions for any field configuration~\cite{RN250}, rendering an unbiased test of DFT performance. Secondly, the quantum phase diagram is generally clear by examining the excitation gap, and phase transitions can be generated by varying the external field~\cite{RN253, RN272}. Thus, a central concern of the energy functional - non-analyticity~\cite{RN263,RN262} - can be studied by purposely sweeping across the quantum critical points or phase boundaries. Lastly, this Ising chain is dual to a fermion chain with p-wave pairing~\cite{RN264}, which undergoes insulator-to-topological-superconductor transitions as the external field varies. Our study therefore also provides insights into DFT of electron systems in this special regime. 

We should mention several pioneering works applying DFT to Heisenberg models~\cite{RN258, RN255, RN256, RN257}. Here, instead of utilizing DFT as an efficient computational tool, our primary motivation is to investigate the capability and structure of DFT itself. In this regard, the quantum Ising chain represents an ideal test bed, which is  ``better controlled than that of the real, necessarily messy situation of actual atoms, molecules and solids''.~\cite{RN252}



\section{Theoretical framework}

\begin{figure}
\centering
    \tikzstyle{formatNoEdge} = []
    \tikzstyle{format}=[rectangle,draw, semithick,fill=white]  
    \tikzstyle{test}=[diamond,aspect=1.5,draw,thin]  
    \tikzstyle{point}=[coordinate,on grid,]  

    \begin{tikzpicture}
    \node[format] (gi){$\{g_i\}$};
    \node[point, left of=gi, node distance = 32mm](labelfigure){};
    \node[point, above of=labelfigure, node distance = 0mm, label=above:{$(a)$}](){};
    \node[point, below of=gi, node distance=9mm] (point0){};
    \draw[-, semithick] (gi)--(point0);
    \node[point, left of=point0, node distance = 19mm] (point1){};
    \node[point, below of=point1, node distance=2mm, label=90:{$\displaystyle \min_{\{\sigma_i^x\}}\ E_{DFT}$}] (minEDFT){};
    \draw[-, semithick] (point0)--(point1);
    \node[point, right of=point0, node distance = 10mm] (point2){};

    \node[formatNoEdge, right of=point2, node distance = 20mm] (Delta){\textcolor{red}{$\Delta$}};
    \draw[->, >=stealth, semithick] (point0)--node[above]{Exact solution}(Delta);
    \node[format, below of=point1, node distance = 9mm] (sigmaxDFT){$\{\sigma_i^{x,\ DFT}\}$};
    \draw[->, >=stealth, semithick] (point1)--(sigmaxDFT);
    \node[format, below of=point2, node distance=9mm] (sigmaexact) {$\{\sigma_i^{x,\ exact}\}$};
    \draw[->, >=stealth, semithick] (point2)--(sigmaexact);
    \draw[<->, draw=red, >=stealth, semithick](sigmaxDFT)--node[below]{\textcolor{red}{$\delta_s$}}(sigmaexact);
    \node[point, right of=sigmaexact, node distance = 26mm](pointBelowDelta){};
    \draw[-, semithick] (sigmaexact)--(pointBelowDelta);
    \node[format, below of=sigmaxDFT, node distance = 13mm](EDFTDFT){$E_{DFT}\left[\sigma_i^{x,\ DFT}\right]$};
    \node[format, below of=sigmaexact, node distance = 13mm](Eexact){$E_{exact}$};
    \node[format, below of=pointBelowDelta, node distance = 13mm](EDFTexact){$E_{DFT}\left[\sigma_i^{x,\ exact}\right]$};
    \draw[->, >=stealth, semithick] (sigmaxDFT)--(EDFTDFT);
    \draw[->, >=stealth, semithick] (pointBelowDelta)--(EDFTexact);
    \draw[<->, draw=red, >=stealth, semithick] (EDFTDFT)--node[below]{\textcolor{red}{$\delta_E^{SC}$}}(Eexact);
    \draw[<->, draw=red, >=stealth, semithick] (Eexact)--node[below]{\textcolor{red}{$\delta_E$}}(EDFTexact);

    \node[point, below of=point2, node distance = 2.5mm] (point3){};
    \node[point, right of=point3, node distance = 12mm] (point4){};
    \node[point, below of=point4, node distance = 5.5mm] (point5){};
    \node[point, below of=point5, node distance = 2mm] (point6){};
    \node[point, below of=point6, node distance = 5mm](point7){};
    \node[point, left of=point7, node distance = 12mm](point8){};
    \draw[-, semithick] (point2)--(point3)--(point4)--(point5);
    \draw[->, >=stealth, semithick] (point6)--(point7)--(point8)--(Eexact);

    \end{tikzpicture}
\includegraphics[width=1.0\columnwidth]{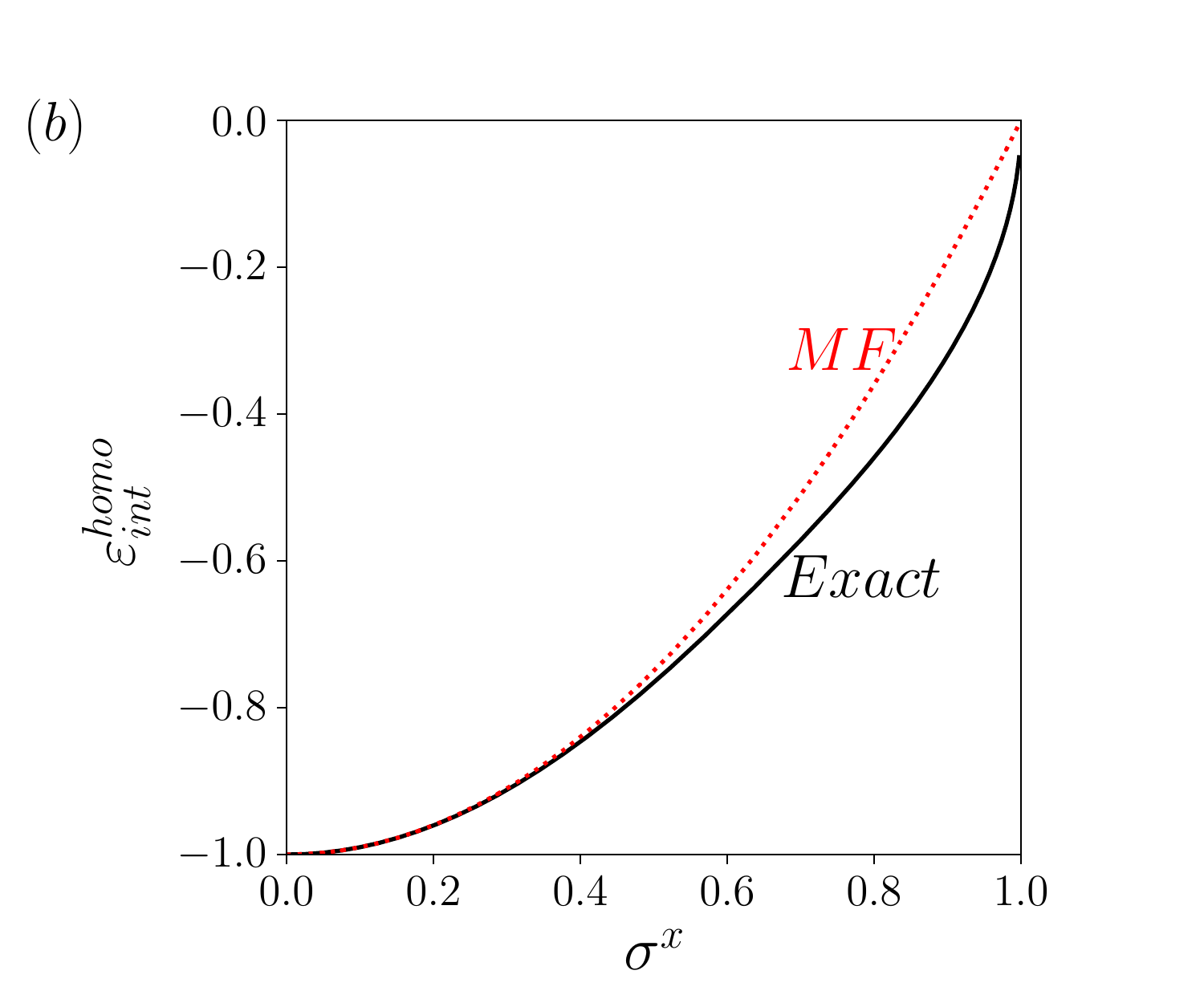}
\caption{(a) Testing procedure. Starting with a given external field $\{g_i\}$, Jordan-Wigner transformation gives the exact excitation gap $\Delta$, ground-state density $\{\sigma_i^{x,\ exact}\}$ and ground-state energy $E_{exact}$ . By substituting exact density into the energy functional $E_{DFT}[\sigma_i^{x,\ exact}]$, the difference between the DFT energy and the exact energy measures the energy error under a given density distribution ($\delta_E$) . In the left branch of the flowchart, we obtain the self-consistent density $\{\sigma_i^{x,\ DFT}\}$ by minimizing the energy functional $E_{DFT}$ under given $\{g_i\}$, and $\delta_s$ measures the density error. The difference between $E_{DFT}[\sigma_i^{x,\ DFT}]$ and $E_{exact}$ measures the energy error under a given field configuration ($\delta_E^{SC}$) . The key quantities used to plot Figs. \ref{fig:summary} and \ref{fig:2D}, including DFT errors ($\delta_s$, $\delta_E$, $\delta_E^{SC}$) and excitation gap ($\Delta$),  are highlighted in red.
 (b) Relation between $\varepsilon_{int}^{homo}$ and  $\sigma^x$. The solid black curve is the exact solution, and the red dashed curve is the MF approximation. The LDA and GGA functionals are constructed to reproduce the exact energies at the homogeneous limit.}\label{fig:1}
\end{figure}
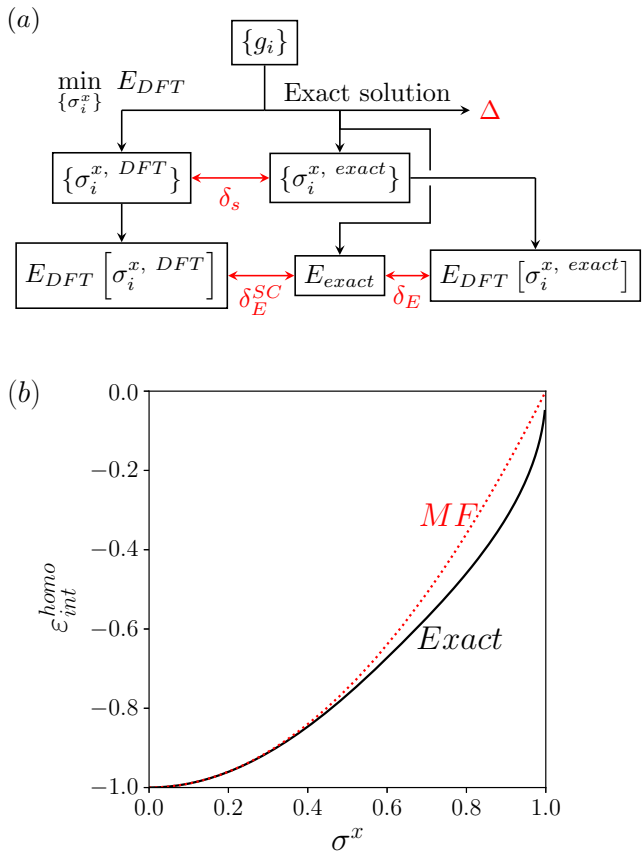

\begin{figure*}
\centering
\includegraphics[width=1.8\columnwidth]{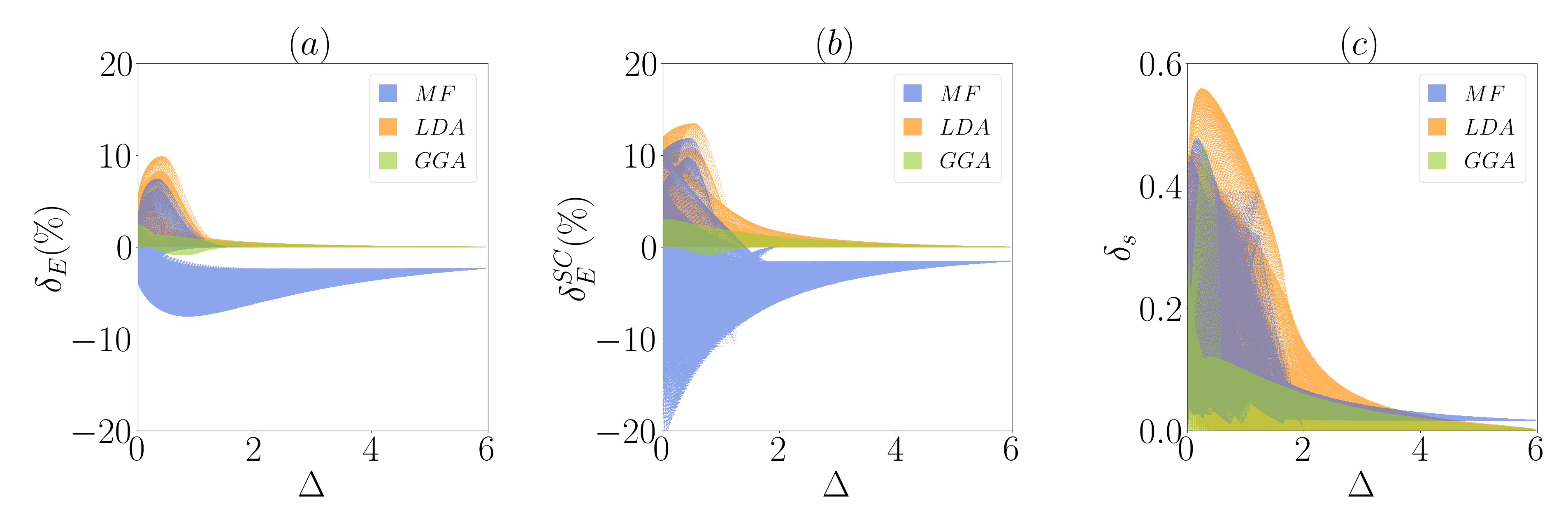}
\caption{Summary of energy and density errors arranged according to the excitation gap $\Delta$ of the system. (a) Energy error of functional with exact density distribution $\delta_E$. (b) Energy error of functional with density distribution minimizes the DFT energy $\delta_E^{SC}$. (c) Density error ($\delta_S$). See Eqs. (\ref{delta_E}) and \ref{delta_s}  for definitions of these errors.}\label{fig:summary}
\end{figure*}

We begin by writing down the Hamiltonian of the quantum Ising chain in line with Eq. (\ref{Hgeneral}):
\begin{eqnarray} \label{quantising}
H_{int}&=&-\sum_{i} \hat\sigma_i^z \hat\sigma_{i+1}^z \\
H_{ext}&=&- \sum_{i} g_i \hat\sigma_i^x.
\end{eqnarray}
The $\hat\sigma_i^{x,z}$ are the familiar Pauli matrices acting on site $i$.

It is well known that Eq. (\ref{quantising}) can be solved by Jordan-Wigner transformation~\cite{RN250}. The mapping between the spin-1/2 degrees of freedom and spinless fermions is chosen to be
\begin{eqnarray}\label{JW}
\hat\sigma_i^x&=&1-2c_i^+c_i \nonumber \\
\hat \sigma_i^z&=&-\prod_{j<i} (1-2c_j^+c_j)(c_i+c_i^+).
\end{eqnarray}
The resulting Hamiltonian is quadratic in the fermion operators:
\begin{eqnarray}\label{isingfermion}
H_{int}&=&-\sum_{i} (c_i^+c_{i+1}+c_{i+1}^+c_i+c_i^+c_{i+1}^++c_{i+1}c_i) \nonumber\\
H_{ext}&=&-\sum_{i} g_i(1-2c_i^+c_i), 
\end{eqnarray}
which can be diagonalized by Bogoliubov transformation. The exact energy then serves as the benchmark of DFT predictions [Fig. \ref{fig:1}(a)]. We also document the excitation gap ($\Delta$) from the exact spectrum, which serves as an important indicator of correlation length and phase transitions. In particular, the continuous phase transition between the two gapped phases along the homogeneous axis is accompanied by a closure of the excitation gap.

From the DFT perspective, the ground-state expectation value $\sigma_i^x\equiv\langle\hat\sigma_i^x\rangle$ can be regarded as the generalized density. DFT asserts that a universal functional $E_{int}[\sigma_i^x]$ for the energy of the intrinsic part can be defined, valid for any external field $\{g_i\}$. For any particular $g_i$, the exact GS energy of the system is the global minimum value of $E_{tot}\equiv E_{int}[\sigma_i^x]+\sum_i g_i \sigma_i^x$, and the density profile $\{\sigma_i^x\}$ that minimizes $E_{tot}$ is the exact ground-state density. This statement can be most conveniently proved by following Levy and Lieb's two-step minimization procedure~\cite{RN267, RN268}.

The existence of phase transitions in the quantum Ising chain immediately suggests that the exact energy functional $E_{int}[\sigma_i^x]$ must be nonanalytic. The hope, however, is that some approximate expressions in analogy to those commonly adopted for interacting electrons are quantitatively useful in at least certain regions of the whole phase diagram, and in turn the triumphs and limitations of these approximations can be better understood.

\subsection{Mean-field approximation}
The lowest-level functional we build is by considering a homogeneous limit $g_i=g$ and by approximating $H_{int}$ in a mean-field (MF) way:
\begin{eqnarray}
H_{int}\approx-\sigma^z\sum_{i} \hat\sigma_i^z,
\end{eqnarray}
in which $\sigma^z\equiv\langle\sigma_i^z\rangle$. The interacting chain is decoupled into individual spins in an effective field $\vec B_{\text{eff}}=(g,0,\sigma^z)$, and the spin ground state can be self-consistently determined to be:
\begin{eqnarray}
\sigma^z=\left\{
\begin{aligned}
&\sqrt{1-g^2},& \quad 0\le g<1 \\
&0, \quad& g\ge1
\end{aligned}
\right. 
\end{eqnarray}
\begin{eqnarray}
\sigma^x=\left\{
\begin{aligned}
&g,& \quad 0\le g<1 \\
&1. \quad& g\ge1
\end{aligned}
\right. 
\end{eqnarray}
Note that for all $g$, the relation $\sigma^z=\sqrt{1-(\sigma^x)^2}$ holds. Accordingly, a compact form of the functional is formulated as:
\begin{eqnarray}\label{MF}
E_{int}^{MF}[\sigma_i^x]=-\sum_{i} [1-(\sigma_i^x)^2].
\end{eqnarray}

This is of course a rather crude approximation, but for a general many-body system without an analytical solution, MF is always a useful starting point. It can be viewed as a counterpart of the Hatree approximation for interacting electrons, in which the quantum effects are largely neglected.

\subsection{Local density approximation}
The merit of local density approximation (LDA) is to assume that the local energy at each point is the same as in a homogeneous system with that density~\cite{RN271}:
\begin{eqnarray}
E_{int}^{LDA}[\sigma_i^x]=\sum_{i} \varepsilon_{int}^{homo}(\sigma_i^x).
\end{eqnarray}
The exact energy density and $\sigma^x$ at any given homogeneous field $g$ can be easily calculated via Eq. (\ref{isingfermion}). Then, the relation between $\varepsilon_{int}^{homo}$ and  $\sigma^x$ can be numerically established. 

We plot $\varepsilon_{int}^{homo}(\sigma^x)$ in Fig. \ref{fig:1} together with the MF approximation: $-[1-(\sigma^x)^2]$. It is clear that the MF energy deviates from the exact one at large $\sigma^x$, where the quantum fluctuation of $\sigma^z$ becomes crucial. 

\subsection{Generalized-gradient approximation}
To move on from the homogeneous limit, we wish to encode the variation of $\sigma_i^x$ into the functional. Formally, we assume that away from the quantum critical points, the energy functional can be expanded into a power series:
\begin{eqnarray}\label{expansion}
E_{int}=A+B\sum_i (\sigma_i^x)^2+C\sum_i (\sigma_i^x)^2(\sigma_{i+1}^x)^2+...
\end{eqnarray}
Note that only even power terms appear, because Eq. (\ref{quantising}) is invariant under a local reflection that reverses the direction of $x$-axis of site $i$. Looking back at Eq. (\ref{MF}), we can see that the MF functional is nothing but the lowest-order Taylor expansion about $\sigma_i^x=0$.  

The third term in Eq. (\ref{expansion}) contains the lowest-order and nearest-neighbor effect of inhomogeneity. To improve LDA by including this term, we should be careful about double counting. Especially, LDA is exact at the homogeneous limit, i.e. $\sigma_i^x=\sigma_{i+1}^x$, and equivalently $\sigma_i^x=-\sigma_{i+1}^x$ with a local axis reflection. We add a correction term to the LDA functional as follows:
\begin{eqnarray}\label{correction}
\Delta E_{int}=C\sum_i [(\sigma_i^x)^2(\sigma_{i+1}^x)^2-(\bar\sigma_i^x)^4-\delta_i^4], 
\end{eqnarray}
in which $\bar\sigma_i^x\equiv(\sigma_i^x+\sigma_{i+1}^x)/2$ and $\delta_i\equiv(\sigma_i^x-\sigma_{i+1}^x)/2$. The last two terms takes care of double counting, which ensure that the correction vanishes at the homogeneous limit. 

Equation (\ref{correction}) can be transformed into $-\frac{C}{8}[(\sigma_i^x)^2-(\sigma_{i+1}^x)^2]^2$. Finally, we arrive at a functional, which we term as generalized-gradient approximation (GGA) by borrowing the terminology from interacting electrons~\cite{RN270}:
\begin{eqnarray}
E_{int}^{GGA}=E_{int}^{LDA}+J_{\text{eff}}\sum_i [(\sigma_i^x)^2-(\sigma_{i+1}^x)^2]^2.
\end{eqnarray}
$J_{\text{eff}}$ is a field-independent parameter fit from the exact energies. For all the numerical calculations below, $J_{\text{eff}}$ is fixed to be 0.35.

\section{Numerical results and discussion}

\begin{figure}
\centering
\includegraphics[width=1.0\columnwidth]{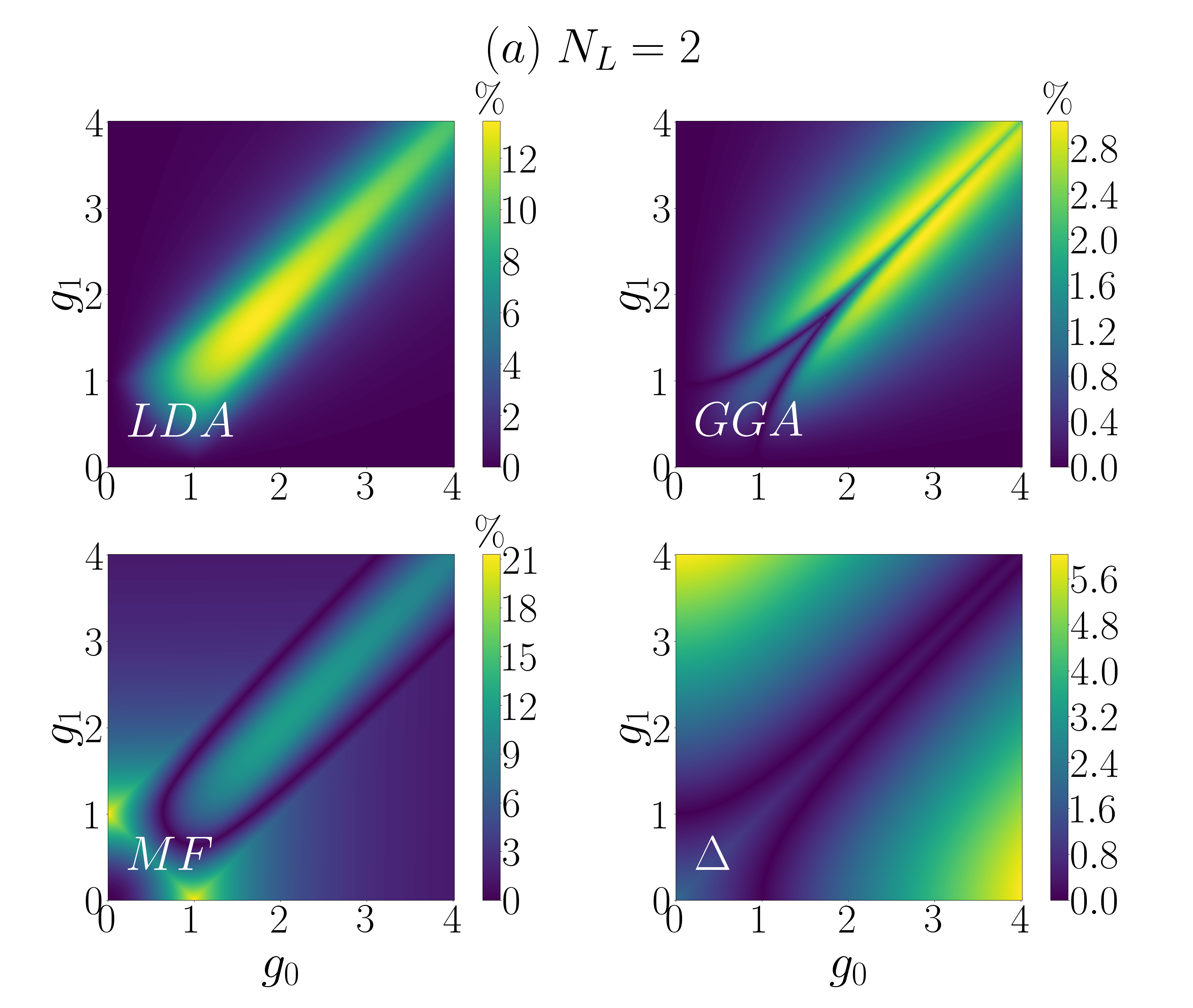}
\includegraphics[width=1.0\columnwidth]{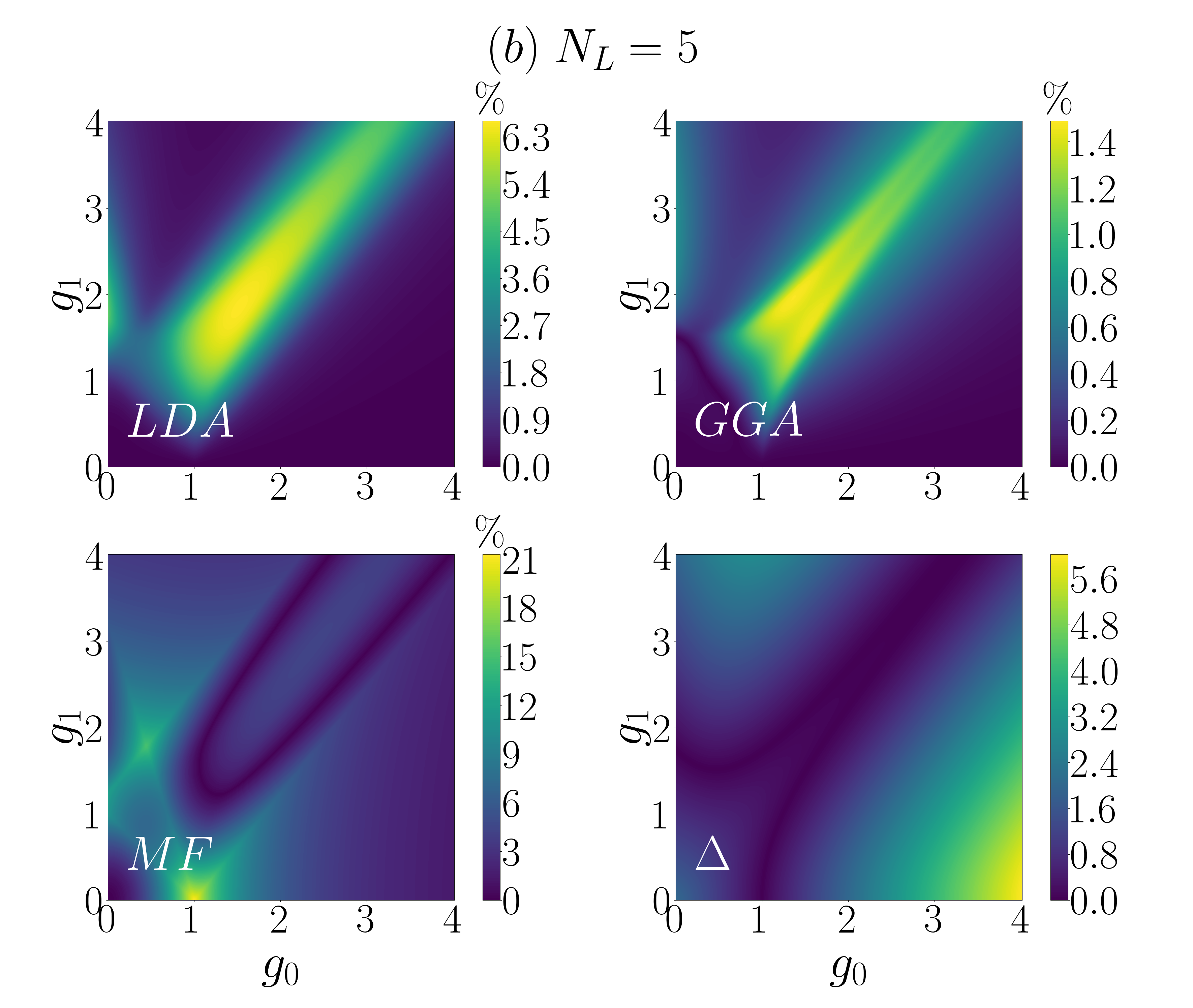}
\includegraphics[width=1.0\columnwidth]{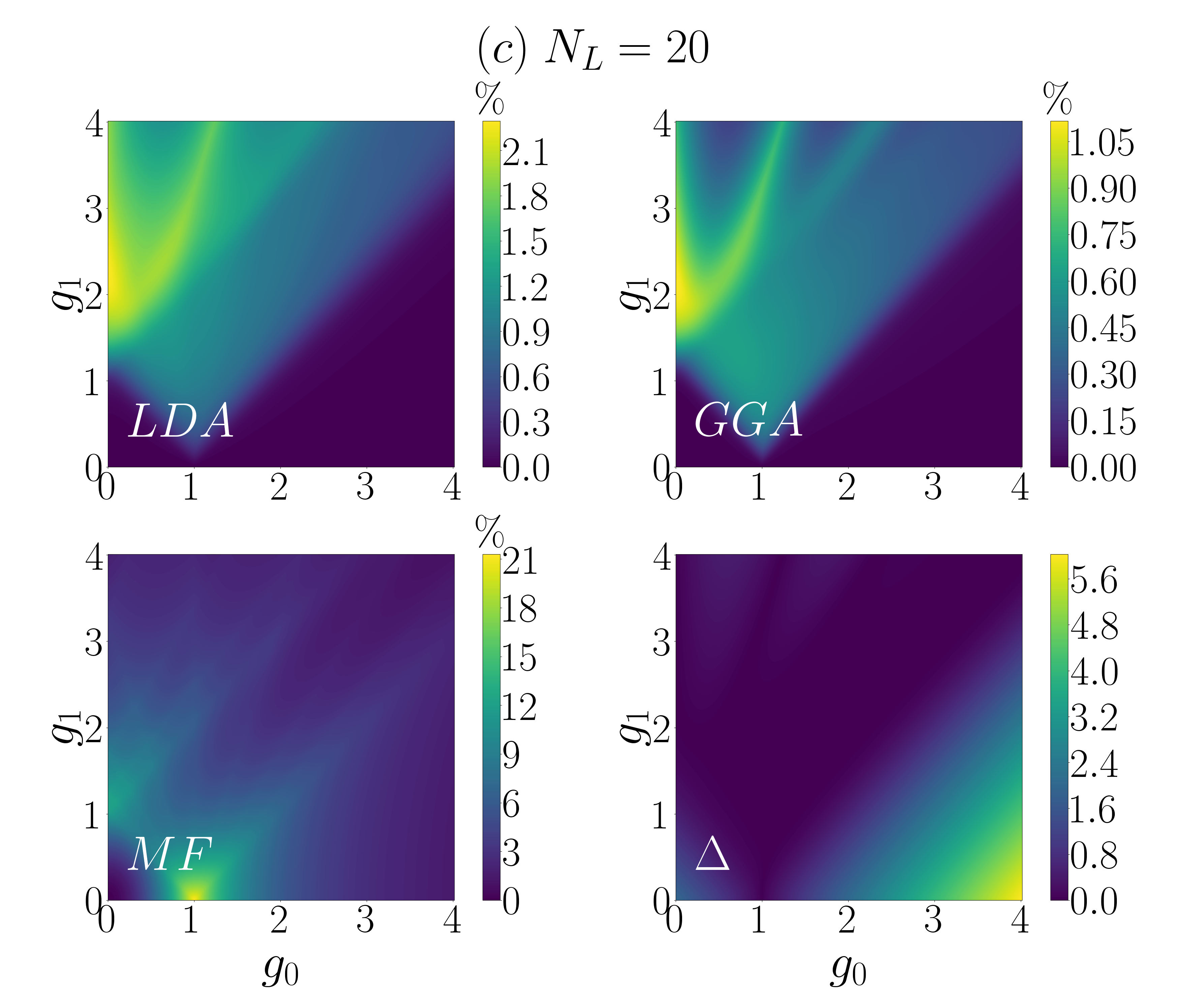}
\caption{Distribution of the energy errors $|\delta_{LDA}^{SC}|$, $|\delta_{GGA}^{SC}|$, $|\delta_{MF}^{SC}|$ and excitation gap $\Delta$ in the $g_0$-$g_1$ plane for (a) $N_L=2$, (b) $N_L=5$ and (c) $N_L=20$.} 
\label{fig:2D}
\end{figure}

To test the performance of these approximated functionals, we consider field configurations in the form of
\begin{eqnarray}
g_i=g_0+g_1 cos (\frac{2\pi i}{N_L}),
\end{eqnarray}
with $g_0$ and $g_1$ ranging from 0 to 4, and $N_L$ ranging from 2 to 20. For each field configuration, the exact solutions are first obtained as the benchmark. 

DFT energies are obtained in two different ways [c.f. Fig.~\ref{fig:1}(a)]: (1) directly plugging in the exact density distribution, $E_{DFT}[\sigma_i^{x,exact}]$; and (2) plugging in the density distribution that minimizes the DFT energy, $E_{DFT}[\sigma_i^{x,DFT}]$. The former energy reflects the precision of the DFT functional when the density is given, and the latter represents the fully self-consistent DFT prediction. The energy error is defined as
\begin{eqnarray}\label{delta_E}
\delta_E=\frac{E_{tot}^{DFT}-E_{tot}^{exact}}{E_{tot}^{exact}}.
\end{eqnarray}
The density error is defined as
\begin{eqnarray}\label{delta_s}
\delta_{s}=\sqrt{\frac{1}{N_L}\sum_{i=1}^{N_L}(\sigma_i^{x,DFT}-\sigma_i^{x,exact})^2}.
\end{eqnarray}

Figure \ref{fig:summary} summarizes the results of 760,000 different field configurations [$200 (g_0)\times200 (g_1)\times19 (N_L)$] arranged according to their gap sizes. Figure \ref{fig:summary}(a) contains the energy errors when the exact density is given. We see that the maximum of $|\delta_E^{GGA}|$ is about $2.5\%$ energy precision, in comparison to about $10\%$ for MF and LDA.  MF shows a long tail of systematic error along the $\Delta$-axis, while both LDA and GGA becomes accurate when the field configuration produces a sufficiently large gap. This is understandable from the fact that correlation length of the Ising chain is inversely correlated with the gap size. When the correlation length is sufficiently short (big gap size), the local or semi-local approximations are well suited. The self-consistent energy errors [Fig. \ref{fig:summary}(b)] are typically larger, due to the discrepancy between the self-consistent density and exact density [Fig. \ref{fig:summary}(c)]. Overall, GGA provides an excellent prediction of the exact energies and densities.

It is natural to speculate that the quantum critical regions are the most difficult to attain a simple density functional, because the correlation length approaches infinity. By plotting the self-consistent energy errors and the excitation gaps in the $g_0$-$g_1$ plane (Fig. \ref{fig:2D}), we observe a clear correspondence between the DFT errors and the gap distribution. At the homogeneous limit ($g_1=0$), both the LDA and GGA functionals are exact. For comparison, notice the peak of the MF error at the homogeneous quantum critical point $g_0=1, g_1=0$. At finite $g_1$, the periodic potential modulates the gap, creating additional quantum critical regions. For a given $g_0$, with increasing $g_1$, there is a certain range in which the LDA and GGA errors barely change.  This optimal performance range is to a large extent bounded by the first collapse of the excitation gap, which signals a quantum phase transition. Although GGA systematically reduces the error, it does not noticeably expand the optimal performance range compared to LDA. In other words, the LDA and GGA error structures are similar (note the changing scales of color bars used in Fig. \ref{fig:2D}).  For MF, the optimal performance range is limited within a smaller region when both $g_0$ and $g_1$ are small. 
When $N_L$ increases, the semi-local approximation becomes better, and the LDA and GGA errors quickly decrease, whereas the MF performance does not improve accordingly.

It is also interesting to notice that $N_L=2$ [Fig. \ref{fig:2D}(a)] represents a special case, for which the LDA and GGA functionals are exact not only along the $g_1=0$ axis but also along the $g_0=0$ axis. The reason is that the staggered field configuration $(-1)^{i}g_1$ is equivalent to the homogeneous one under the local $\hat x\rightarrow-\hat x$ reflection symmetry. Consequently, the error structure is symmetric with respect to $g_0=g_1$, and the DFT functionals perform equally well when $g_1\gg g_0$. The lesson is that by taking advatange of certain symmetries of the system, it is possible to expand the application of DFT to a wider range of the parameter space.

In the fermion dual representation [Eq. (\ref{JW})], $E_{int}[\sigma_i^x]\rightarrow E_{int}[n_i]$, giving the more familiar form of DFT in terms of the fermion local density. It is peculiar to notice that our $E_{int}[\sigma_i^x]$ directly includes the kinetic energy without invoking the Kohn-Sham ansatz~\cite{RN271}, whereas the kinetic energy is typically known to lead to nonlocality and nonanalyticity in the functional~\cite{RN262}. We consider that two special aspects of the quantum Ising chain makes such a particularly simple form of energy functional possible.

(i) In the neighborhood of the homogeneous axis, the fermions fall either to an insulating or a fully-gapped superconducting phase, which guarantees an exponential decay of the correlation function, except for the quantum critical point. 

(ii) The definition domain of the energy functional is $(-1,1)$ in terms of $\sigma_i^x$ or $(0,1)$ in terms of $n_i$. Staying within a specific phase, the energy functional is smooth, and the high-order contribution of the Taylor expansion [Eq. (\ref{correction})] is expected to vanish rapidly. 

\newcommand\crossmark[1][]{%
  \tikz[scale=0.25,#1]{
    \fill(0,0)--(0.1,0) .. controls (0.5,0.4) .. (1,0.7)--(0.9,0.7) ..  controls (0.5,0.5) ..(0,0.1) --cycle;
    \fill(1,0.1)--(0.9,0.1) .. controls (0.5,0.3) .. (0,0.7)--(0.1,0.7) .. controls (0.5,0.4) ..(1,0.2) --cycle;
  }%
}

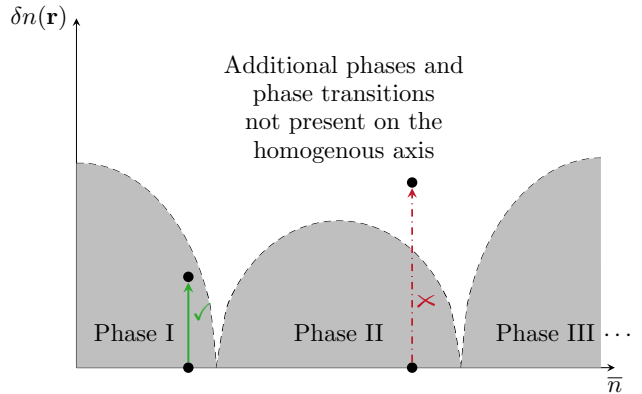
\begin{figure}[t]
\centering
    \resizebox{\columnwidth}{!}{%
    \begin{tikzpicture}
        \draw[->, >=stealth] (0, 0) -- (7.7,0) node[below]{$\overline{n}$};
        \draw[->, >=stealth] (0, 0) -- (0, 5) node[left]{$\delta n(\mathbf{r})$};

        \draw[densely dashed,domain=0:2, smooth, variable=\x, black] plot ({\x}, {1.5 * sqrt(abs((\x+1.9)*(2 -\x)))});
        \fill [lightgray, domain=0:2, variable=\x]
            (0, 0)
            -- plot ({\x}, {1.5 * sqrt(abs((\x+1.9)*(2 -\x)))})
            -- (2, 0)
            -- cycle;
        \draw[densely dashed,domain=2:5.5, smooth, variable=\x, black] plot ({\x}, {1.2 * sqrt(abs((\x - 2)*(5.5 -\x)))});
        \fill [lightgray, domain=2:5.5, variable=\x]
            (2, 0)
            -- plot ({\x}, {1.2 * sqrt(abs((\x - 2)*(5.5 -\x)))})
            -- (5.5, 0)
            -- cycle;
        \draw[densely dashed,domain=5.5:7.5, smooth, variable=\x, black] plot ({\x}, {1.5 * sqrt(abs((\x - 5.5)*(9.5 -\x)))});
        \fill [lightgray, domain=5.5:7.5, variable=\x]
            (5.5, 0)
            -- plot ({\x}, {1.5 * sqrt(abs((\x - 5.5)*(9.5 -\x)))})
            -- (7.5, 0)
            -- cycle;
        \node[] at (0.825, 0.5) {Phase I};
        \node[] at (3.75, 0.5) {Phase II};
        \node[] at (7, 0.5) {Phase III $\cdots$};
        \node[circle,fill,inner sep=1.5pt] (p0) at (1.6, 0){};
        \node[circle,fill,inner sep=1.5pt] (p1) at (1.6, 1.3){};
        \draw[->, >=stealth, thick, color={rgb:red,50;green,200;blue,50}] (p0)--(p1);
        \node[] at (1.8, 0.8) {\textcolor{rgb:red,50;green,200;blue,50}{\checkmark}};
        \node[circle,fill,inner sep=1.5pt] (p2) at (4.8, 0){};
        \node[circle,fill,inner sep=1.5pt] (p3) at (4.8, 2.65){};
        \draw[->, >=stealth, dashdotted, line width=0.7pt, color={rgb:red,240;green,30;blue,50}] (p2)--(p3);
        \node[] at (5, 0.95) {\textcolor{rgb:red,240;green,30;blue,50}{\crossmark}};
        \node[align=center] at (3.825, 3.7) {
            Additional phases and \\phase transitions \\
            not present on the\\ homogenous axis
            };
    \end{tikzpicture}
    }%
\caption{Schematic quantum phase diagram in the density parameter space. The green arrow marks an inhomogeneous point, which can be analytically connected to the homogeneous axis.  The red arrow marks a nonanalytic path intersected by the phase boundary. }\label{fig:schematic}
\end{figure}

\section{Conclusions}

As a summary, we schematically draw a general quantum phase diagram in Fig. \ref{fig:schematic}. It is understood that the LDA and GGA functionals are built upon the homogeneous limit, and we assume that along the horizontal homogeneous axis, the system can have multiple quantum phases as the average density $\bar n$ varies. These phases are expected to be stable under a finite inhomogeneous modulation $\delta n(r)$, but when the modulation is strong, phase transitions may occur and the system enters a new regime, which is not analytically connected to the homogeneous axis. Our test on the quantum Ising chain suggests that the application of LDA or GGA is largely bounded by the boundaries of the homogeneous phases. Within these boundaries, a properly parameterized semi-local functional is possible to achieve a high level of accuracy, but the challenge is how to extend it to a strongly inhomogeneous point (red arrow in Fig. \ref{fig:schematic}). We believe that this conclusion is generally relevant to interacting electrons and other many-body systems. 




\section*{Acknowledgements}
We thank Hui Zhai and Zheng-Yu Weng for helpful discussion. This work is supported by Tsinghua University Initiative Scientific Research Program,  NSFC (Grant Nos. 11774196, 51788104), and the Ministry of Science and Technology of China (Grant No. 2016YFA0301001)

J. M and H. T contribute equally to this work.

\bibliography{ref.bib}
\end{document}